\documentclass[aps,prb,twocolumn,groupedaddress,showpacs]{revtex4}

\usepackage{graphicx}
\usepackage{dcolumn}
\usepackage{bm}
\begin{document}

\title{Spin reorientation and in-plane magnetoresistance of lightly
doped La$_{2-x}$Sr$_{x}$CuO$_{4}$ in magnetic fields up to 55 T}

\author{S. Ono}
\author{Seiki Komiya}
\author{A. N. Lavrov}
\altaffiliation{Present address: Institute of Inorganic Chemistry,
Novosibirsk 630090, Russia}
\author{Yoichi Ando}
\email[]{ando@criepi.denken.or.jp} 
\affiliation{Central Research Institute of Electric Power Industry, 
Komae, Tokyo 201-8511, Japan}

\author{F. F. Balakirev}
\author{J. B. Betts}
\author{G. S. Boebinger}
\altaffiliation{Present address: National High Magnetic Field Laboratory,
Tallahassee, FL 32310, USA}
\affiliation{NHMFL, Los Alamos National Laboratory, Los Alamos,
New Mexico 87545, USA}

\date{\today}

\begin{abstract}

The magnetoresistance (MR) in the in-plane resistivity is measured in
magnetic fields up to 55 T in lightly doped La$_{2-x}$Sr$_{x}$CuO$_{4}$
in the N\'eel state ($x$ = 0.01) and in the spin-glass state ($x$ =
0.03) using high-quality untwinned single crystals. In both cases, a
large negative MR is observed to appear when the magnetic order is
established. For $x$ = 0.01, it is found that the MR is indicative of a
one-step transition into a high-field weak-ferromagnetic state at around
20 T when the magnetic field is applied from the spin easy axis ($b$
axis), which means that there is no spin-flop transition in the N\'eel
state of this material; this is contrary to a previous report, but is
natural in light of the peculiar in-plane magnetic susceptibility
anisotropy recently found in this system. In the spin-glass state, we
observe that the large (up to $\sim$20\%) negative MR saturates at
around 40 T, and this MR is found to be essentially isotropic when the
magnetic field is rotated within the $ab$ plane. Our data show that the
large negative MR is inherent to LSCO in a magnetically ordered state,
in which the weak-ferromagnetic (WF) moment becomes well-defined; we
discuss that the observed MR is essentially due to the reorientation of
the WF moments towards the magnetic field direction both in the N\'eel
state and in the spin-glass state.

\end{abstract}


\maketitle

\section{INTRODUCTION}

In the lightly doped regime of the high-$T_c$ cuprates, there is an
intriguing dichotomy regarding the coupling between charge carriers and
the background spins: On one hand, the long-range N\'eel order is
quickly suppressed with doped holes, and only 2\% of hole doping is
sufficient to kill the N\'eel ordering, \cite{Kastner} indicating that
the doped holes are strongly coupled to the spin subsystem; on the other
hand, a metallic transport with a mobility comparable to that at optimum
doping is established with only 1\% of hole doping and this metallic
transport is completely insensitive to the onset of the N\'eel ordering,
\cite{mobility} indicating that the doped holes and the spin subsystem
are rather decoupled. Even more intriguingly, the dichotomy is found not
only in the hole-doped cuprates but also in the electron-doped cuprates,
where the N\'eel temperature is rather insensitive to the electron
doping \cite{Keimer} (suggestive of a spin-charge decoupling), and yet a
subtle change in the spin arrangement upon the spin-flop transition in
magnetic fields causes a large change in the resistivity
\cite{LavrovPLCCO} (indicative of a strong spin-charge coupling).
Therefore, such dichotomy appears to be a ubiquitous feature in the
cuprates (though there is an electron-hole asymmetry in its nature) and
it is of high importance to clarify the cause of this puzzle to
understand the peculiar electronic states in the cuprates. As has been
pointed out repeatedly in the literature,
\cite{mobility,Zaanen,Kivelson,Machida,Cho,Mueller,anisotropy} some form
of electron self-organization and a resulting nanoscale phase separation
is probably the key to solve this puzzle, but a comprehensive picture to
understand the interplay between the charge carriers and the background
spins, which is clearly at the heart of the physics of the cuprates,
remains to be developed.

The magnetoresistance (MR) in the lightly doped cuprates has been a
useful probe of the peculiar coupling between the charge carriers and
the spin subsystem. For example, Thio {\it et al.} discovered
\cite{ThioWF} a large change in the resistance of a lightly oxygen-doped
La$_{2}$CuO$_{4+y}$ (LCO) sample at the weak-ferromagnetic (WF)
transition that occurs at around 4 T when the magnetic-field is applied
along the $c$ axis. They interpreted this phenomenon in terms of a sort
of ``spin-valve" effect, which controls the hopping probability
depending on the ferromagnetic/antiferromagnetic arrangements of the WF
moments between the neighboring CuO$_2$ planes and thereby changing the
$c$-axis resistivity $\rho_c$. \cite{ThioFlop} Remember, the WF moment
in LCO [or in antiferromagnetic La$_{2-x}$Sr$_{x}$CuO$_{4}$ (LSCO)]
arises from a slight canting of the Cu spins out of the CuO$_2$ planes
due to the antisymmetric exchange coming from the Dzyaloshinskii-Moriya
(DM) interaction \cite{Kastner,ThioWF,Thio94}; when there is a static
in-plane antiferromagnetic order, because of the spin canting, each
CuO$_2$ plane develops a weak $c$-axis ferromagnetic moment, which
orders antiferromagnetically along the $c$ axis in the three-dimensional
N\'eel state but can be aligned ferromagnetically upon WF transition
when a sufficiently strong $c$-axis magnetic field is applied. The MR in
lightly doped LSCO was recently revisited by Ando, Lavrov, and Komiya
(ALK) \cite{AndoMR} and they found that not only the $\rho_c$ but also
the in-plane resistivity ($\rho_{ab}$) shows a large change upon the WF
transition, which cannot be explained in terms of the spin-valve
scenario. ALK therefore proposed an alternative model in which they
asserted that the in-plane charge transport occurs primarily through a
network of antiferromagnetic domain boundaries (that can be viewed as
nematic charge stripes) and that a change in the nature of the boundary
from antiphase to in-phase (which necessarily happens upon the WF
transition) should be the cause of the large MR in $\rho_{ab}$.
\cite{AndoMR}

Although the nature of the WF transition for $H \parallel c$ is by now
well understood, \cite{AndoMR} the case for $H \perp c$ is rather
controversial. Based on MR measurements of flux-grown LCO crystals for
$H \parallel a,b$ up to 23 T, in 1990 Thio {\it et al.} asserted
\cite{ThioFlop} that when the magnetic field is applied along the spin
easy axis ($b$ axis), the spin reorientation takes place in three steps:
first, as the WF moment is drawn towards the magnetic-field direction
($b$ axis), the staggered moment is gradually reoriented towards the $c$
axis within the $bc$ plane; second, an ordinary spin-flop transition
takes place and the staggered moment flops into the $ac$ plane,
perpendicular to the applied field; third, as the WF moment is kept
drawn towards the magnetic-field direction, the staggered moment is
gradually reoriented towards the $c$ axis within the $ac$ plane. In the
final high-field state, the staggered moment is essentially along the
$c$ axis and the WF moments are ferromagnetically aligned along the $b$
axis (we call this state a high-field WF state). However, very recently,
Gozar {\it et al.} \cite{Gozar} measured the long-wavelength magnetic
excitations in untwinned single crystals of undoped LCO and lightly
doped LSCO using Raman spectroscopy and argued that there should be no
intermediate ``spin-flopped" state where the staggered moments are in
the $ac$ plane; in other words, the spin reorientation in this system
takes place in just one step, namely, a gradual rotation of the
staggered moment towards the $c$ axis within the $bc$ plane until it
reaches the $c$ axis to realize the high-field WF state. (It is useful
to note that Gozar {\it et al.} found a new field-induced magnetically
ordered state above the N\'eel temperature $T_N$ for $H \parallel b$,
which is essentially the same as the high-field WF state below
$T_N$.\cite{Gozar})

Therefore, it is desirable to conduct careful MR measurements on a
state-of-the-art single crystals of lightly doped LSCO up to a high
magnetic field and clarify whether the transition into the high-field WF
state is achieved in one step or in multiple steps when the magnetic
field is applied along the $b$ axis. Also, since the magnetic-field
induced spin reorientation in the spin-glass regime \cite{Kastner} of
LSCO ($0.02 \le x \le 0.05$) has not been studied before, it would be
useful to measure the high-field MR in the spin-glass regime as well:
for example, recent Raman spectroscopy experiment by Gozar {\it et al.}
\cite{Gozar} mentioned above did not find any feature that can be
associated with the field-induced magnetically ordered state for $x$ =
0.02 or 0.03, so it is intriguing to see what is observed in the MR in
those samples.

In this work, we measure MR in the in-plane resistivity of lightly doped
LSCO at $x$ = 0.01 and 0.03 in pulsed magnetic fields up to 55 T. We
find that at $x$ = 0.01 there is indeed no intermediate spin-flop
transition for $H \parallel b$ before the high-field WF state is
achieved at the critical field $H_b$ of $\sim$20 T. Intriguingly, while
a large negative MR is observed below $H_b$, the MR above $H_b$ (in the
high-field WF state) is found to be positive; this seems to suggest that
in cuprates an antiferromagnetic alignment of the background spins gives
better charge conduction than a ferromagnetic (spin polarized)
alignment, which is contrary to the common wisdom for metals. The MR
data for $x$ = 0.03 suggest that the charges are strongly coupled to the
spin subsystem only at sufficiently low temperature where the spins
freeze into a spin-glass state; in this case, a saturation of the
negative MR occurs at $\sim$40 T, which is roughly twice as much as that
for $x$ = 0.01. What is unusual for $x$ = 0.03 is that both the
saturation field and the magnitude of the MR appear to be almost
isotropic with respect to the rotation of the magnetic field in the $ab$
plane; we discuss that this unexpected isotropy can be understood to be
a result of a short antiferromagnetic correlation length in the
spin-glass state, and that the mechanism to produce large negative MR at
$x$ = 0.01 and 0.03 is essentially the same.

\section{EXPERIMENTS}

The high-quality LSCO single crystals are grown by the traveling-solvent
floating-zone technique. \cite{Komiya} After they are cut and polished
into a rectangular platelet shape of 1.2$\times$0.6$\times$0.15 mm$^3$
suitable for the $a$-axis resistivity ($\rho_a$) measurements (the
longest edge is parallel to the $a$ axis within an error of less than
1$^{\circ}$), the samples are carefully annealed at 500$^{\circ}$C in
flowing pure Ar to remove the excess oxygen, and then detwinned at
250$^{\circ}$C with a uniaxial pressure. \cite{sus} The samples are
confirmed to be nearly 100\% detwinned by the x-ray diffraction (XRD)
analysis. The MR is measured in pulsed magnetic fields up to 55 T using
a high-frequency ($\sim$100 kHz) lock-in technique with a four-probe
method. \cite{logT,MI} The eddy current heating \cite{logT} is confirmed
to be not adversely affecting the data shown here; however, in the case
of the $x$ = 0.01 sample, the MR data below 20 K are found to be
unreliable, because the resistivity of this sample diverges almost
exponentially at low temperature (see Fig. 1), which causes the
resistivity value to be extremely sensitive to even a slight change in
temperature due to a minor eddy-current heating.

\begin{figure}
\includegraphics[width=8cm]{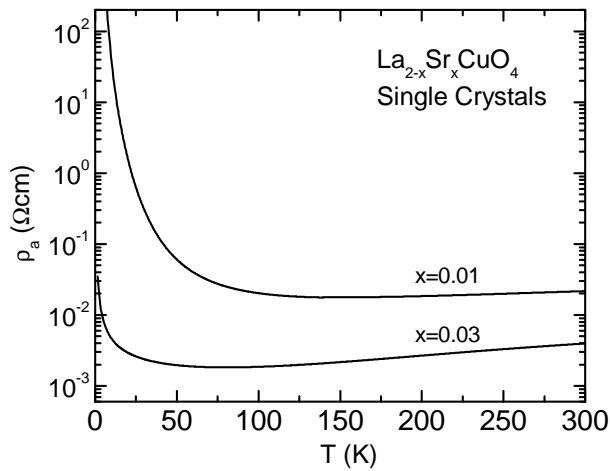}
\caption{Temperature dependences of $\rho_a$ for $x$ = 0.01 and 0.03,
measured on the untwinned samples used in this study.}
\end{figure}

Figure 1 shows the temperature dependences of $\rho_a$ measured on the
samples used in the present study. Note the difference in the
low-temperature behavior for the two samples: The $x$ = 0.01 sample,
which is in the N\'eel state below $\sim$230 K, shows a steep
resistivity divergence at low temperatures, while the $x$ = 0.03 sample
shows comparatively modest resistivity divergence.

\section{RESULTS AND DISCUSSIONS}

\subsection{N\'eel State ($x$ = 0.01)}

\begin{figure}
\includegraphics[width=8.5cm]{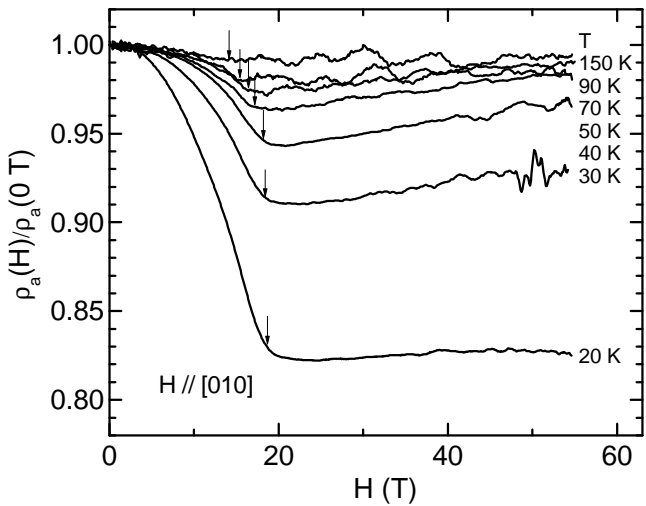}
\caption{Magnetoresistance in $\rho_a$ with $H \parallel b$ for the
$x$ = 0.01 sample at various temperatures. Arrows mark the transition
field into the high-field WF state, $H_b$.}
\end{figure}

The MR in $\rho_a$ with $H \parallel b$ for $x$ = 0.01 is shown in Fig.
2. Although the data are noisy at high temperatures, one can see that
the low-field negative MR at 150 K tends to saturate at around 13 T,
which is consistent with our previous data measured in dc magnetic field
(see the 160 K data in Fig. 2 of Ref. \onlinecite{AndoMR}), giving
confidence in the present pulsed magnetic field measurements. It should
be noted that, according to the previous study, \cite{AndoMR} only the
$b$-axis component of the in-plane magnetic field is responsible for the
low-field negative MR. As the temperature is lowered, the absolute value
of the low-field negative MR grows [Fig. 3(a)] and the critical field
moves to higher field [Fig. 3(b)], both of which are consistent with the
previous reports. \cite{ThioFlop,AndoMR} However, there are two features
that are different from the old data of Thio {\it et al.}
\cite{ThioFlop}: (i) the MR above $H_b$ is positive, and (ii) there is
no kink in the MR data below $H_b$. The latter indicates that there is
no intermediate spin-flop transition proposed by Thio {\it et al.} The
absence of the spin-flop transition for $H \parallel b$ is actually
natural in light of the recent magnetic susceptibility data \cite{sus}
measured on an untwinned single crystal of LSCO at $x$ = 0.01, which
showed that $\chi_a$ is always smaller than $\chi_b$ at temperatures
below 300 K, even though the $b$ axis is the spin easy axis; remember,
in ordinary antiferromagnets in the N\'eel state, the perpendicular
susceptibility $\chi_{\perp}$ is larger than the longitudinal
susceptibility $\chi_{\parallel}$, which provides the source of the
energy gain in the spin-flop transition. Therefore, the unusual
anisotropy in the magnetic susceptibility (which is probably due to the
peculiar role of the DM interaction \cite{sus}) makes the occurrence of
the ordinary spin-flop transition unlikely in this system.

\begin{figure}
\includegraphics[width=7cm]{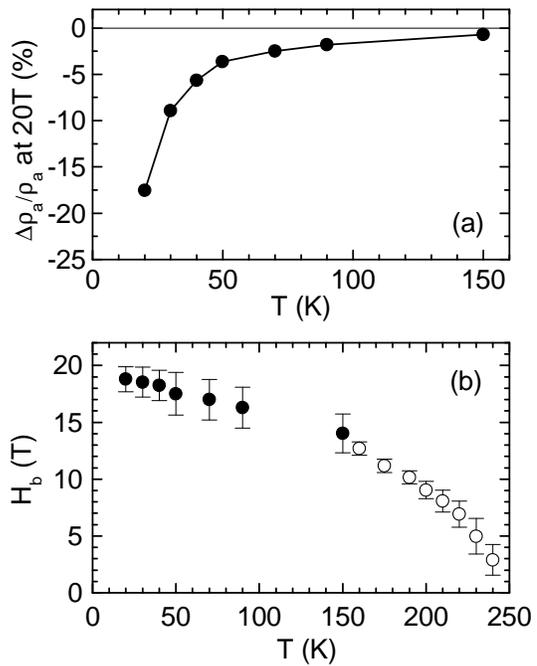}
\caption{(a) Temperature dependence of the magnetoresistance in
$\rho_a$ with $H \parallel b$ for the $x$ = 0.01 at 20 T. 
(b) Temperature dependence of the transition
field into the high-field WF state, $H_b$, extracted from the present
pulsed magnetic field data (solid circles) and the previous dc magnetic
field data \cite{AndoMR} (open circles).}
\end{figure}

Another point to be elaborated on is the positive MR observed above
$H_b$ in Fig. 2. This MR is not likely to be the ordinary positive
(orbital) MR of a metal, because (i) the system is in the strongly
localized (hopping conduction) regime where the electron motion is not
coherent \cite{note} and (ii) the magnetic-field ($H$) dependence of the
MR (particularly at 30 -- 90 K) is not $\sim H^2$ but is rather linear
in $H$. Thus, the MR above $H_b$ is most likely caused by a change in
the spin subsystem. If so, the positive MR is rather unexpected, because
above $H_b$ the main moment is gradually polarized at the expense of the
staggered component of the moment [all the spins are expected to be
ferromagnetically aligned (completely polarized) when the Zeeman energy
exceeds the exchange energy (order of 100 meV in cuprates) and the
thermal energy]. Remember, it is a common sense that the ferromagnetic
arrangement of the spins promotes the charge motion while the
antiferromagnetic arrangement tends to localize the charges; therefore,
according to the common wisdom, the positive MR above $H_b$ in LSCO is
unusual in that the charge mobility tends to become worse when the
system is approaching a spin-polarized state. However, the argument of
whether the ferromagnetic/antiferromagnetic arrangement promotes/hinders
the charge motion is based on a picture that the charges are essentially
uniformly distributed and the physics is determined by the motion of a
single charge in the magnetic background. If, on the other hand, the
charges are phase segregated from the magnetic subsystem and form a
self-organized network of ``rivers of charges", the above consideration
does not apply. In fact, in our series of works on the lightly doped
cuprates, we have shown that various peculiar properties in the lightly
doped regime can be best understood in terms of the electron
self-organization picture.
\cite{mobility,anisotropy,AndoMR,sus,stripe,nonlinear} Therefore, the
positive MR observed above $H_b$ is another indirect support to the
notion that charges and spins are spatially decoupled in lightly doped
LSCO, but the exact reason why the MR is positive in the high-field WF
state (where the ``rivers of charges" must constitute in-phase magnetic
domain boundaries \cite{AndoMR}) needs to be clarified by future
studies.

\subsection{Magnetic Shape Memory Effect}

We note that one should always be careful about the ``magnetic shape
memory effect" \cite{shape} in lightly doped LSCO crystals when making
measurements under high magnetic fields. This effect causes a swapping
of the orthorhombic $a$ and $b$ axes in a {\it fixed} sample, where the
$b$ axis tends to become parallel to the applied in-plane magnetic
field. Therefore, no axes swapping takes place in untwinned crystals
when the magnetic field is within $\pm 45^{\circ}$ of the $b$ axis
direction. However, when a high magnetic field (usually in excess of 10
T) is applied within $\pm 45^{\circ}$ of the $a$ axis, new twin
boundaries are created and initially untwinned crystals become twinned.
In our previous work using dc magnetic fields up to 14 T, \cite{shape}
we observed that this phenomenon takes place for $x$ = 0.01 only near
room temperature where the thermal fluctuations help the axes swapping,
but in the present 55 T experiment we found that this phenomenon can
happen even at low temperatures when the applied magnetic field is
sufficiently strong. In fact, after the $H \parallel b$ measurements are
finished, we tried to measure the MR for $H \parallel a$ in our $x$ =
0.01 sample at low temperatures (below 70 K), but the sample was
re-twinned during this experiment, making the data to be not very
meaningful (this is why only the data for $H \parallel b$ are shown
here).

\subsection{Spin-Glass State ($x$ = 0.03)}

\begin{figure}
\includegraphics[width=8.5cm]{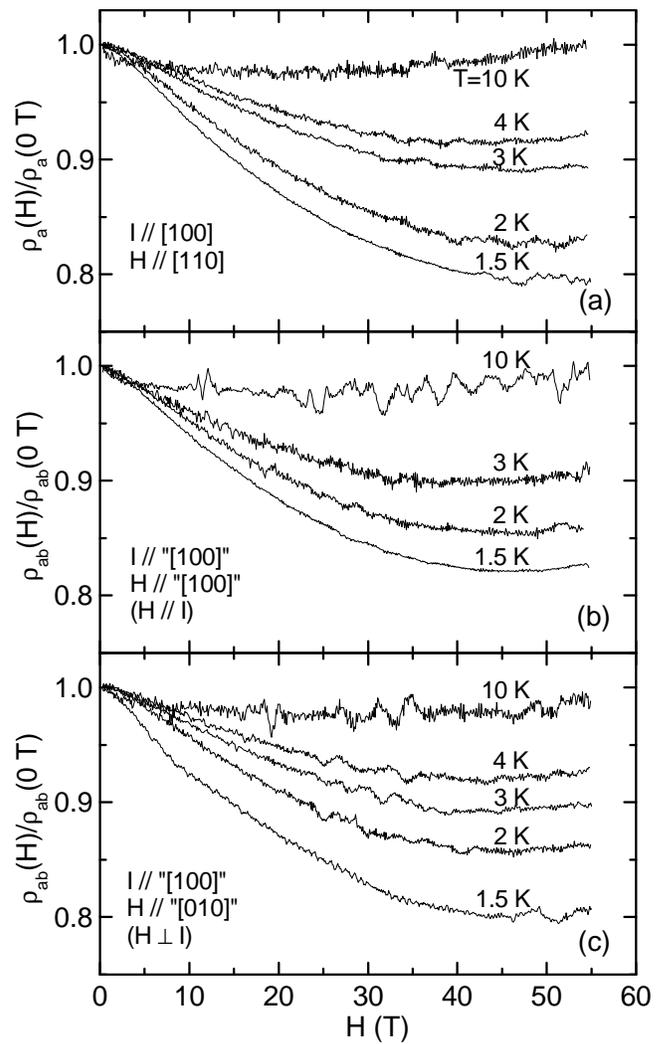}
\caption{Magnetoresistance of the $x$ = 0.03 sample for three
different configurations of the in-plane magnetic field.}
\end{figure}

Figure 4 shows the MR for $x$ = 0.03, where the magnetic field is
applied from three different in-plane directions, $H \parallel [110]$
(diagonal to the orthorhombic axes and thus is parallel to the Cu-O-Cu
bond direction), $H \parallel [100]$ (orthorhombic $a$ axis), and $H
\parallel [010]$ (orthorhombic $b$ axis); we note that the measurements
are done in this order. This $x$ = 0.03 sample was initially 100\%
detwinned and was prepared so that the current flows along the $a$ axis
($I \parallel [100]$), and we did not expect the magnetic shape memory
effect to occur at this composition; however, to our surprise, we found
that the sample was retwinned after the measurements, with about 45\% of
misoriented domains. Since the axes swapping never occurs for $H
\parallel [110]$ (first set of the data), it must have occurred during
the measurement for $H \parallel [100]$ (second set of the data).
Therefore, we should consider that the data for $H \parallel [100]$ and
$H \parallel [010]$ [Figs. 4(b) and 4(c)] are actually on a twinned
sample; this is why the current directions and the magnetic field
directions are denoted using double-quotation marks in Figs. 4(b) and
4(c). Nevertheless, the data shown in Fig. 4 are useful, because they
show that the MR is essentially isotropic at $x$ =0.03: If the MR were
intrinsically anisotropic and the nature of the anisotropy were similar
to that for $x$ = 0.01 ({\it i.e.}, only the $b$-axis component of the
in-plane magnetic field is effective in causing the negative MR), the
characteristic magnetic field scale for the MR would be $\sqrt{2}$ times
larger for $H \parallel [110]$ compared to that for $H \parallel [010]$
even in a twinned sample. In Figs. 4(a)-4(c), the minimum in MR at each
temperature occurs at essentially the same magnetic field for the three
configurations, indicating that both the $a$- and $b$-axis components of
the magnetic field are equally effective in causing the MR. Note that
not only the characteristic magnetic field scale but also the magnitude
of the MR is essentially isotropic, growing up to $\sim$20\% at 1.5 K
for all three configurations.

It is useful to note that the MR for $x$ = 0.03 shown in Fig. 4 remains
small down to 10 K, but suddenly grows at lower temperatures. Thus, the
growth of the MR is clearly related to the spin-glass order, which is
established below 6--10 K for $x$ = 0.03. \cite{sus,Wakimoto}
Intriguingly, in the magnetic susceptibility measurement of an untwinned
single crystal with $x$ = 0.03 by Lavrov {\it et al.}, \cite{sus} it was
found that the spin-glass ordering temperature $T_g$ is anisotropic,
meaning that the spin-glass order parameter vanishes upon heating at
different temperatures depending on the direction of the applied field
to measure the magnetic susceptibility. Obviously, this anisotropy in
$T_g$ is related to the anisotropy in the magnetic susceptibility itself
that is indicative of the staggered moment to be confined in the $bc$
plane, as is the case with $x$ = 0.01; this suggests that the direction
of the spins are not random in the ``spin-glass" phase of LSCO, a
departure from the conventional picture of the spin glass or the cluster
spin glass. Since the local spin structure as suggested by the magnetic
susceptibility anisotropy at $x$ = 0.03 appears to be similar to that at
$x$ = 0.01, it is natural to expect that the negative MR observed for
$x$ = 0.03 is caused by essentially the same mechanism as that for $x$ =
0.01, namely, the reorientation of the WF moments towards the
magnetic-field direction. The fact that the characteristic magnetic
field for the saturation of the MR for $x$ = 0.03 ($\sim$40 T) is larger
than that for $x$ = 0.01 ($\sim$20 T) is also consistent with this
interpretation, because in the spin-glass phase the spin correlation
length is rather short (4--10 nm depending on the in-plane direction)
\cite{Matsuda} and thus the magnitude of the WF moment (which is given
by an integration of the canted moments over the antiferromagnetically
correlated area in the CuO$_2$ planes) becomes accordingly small.

One might think that the isotropic nature of the MR for $x$ = 0.03
evident in Fig. 4 is not very consistent with the interpretation that
the WF moment (that is confined in the $bc$ plane) is responsible for
the MR. However, one should keep in mind that the antiferromagnetically
correlated region in the spin-glass regime is rather small, which makes
it easier for the applied magnetic field, with the help of the thermal
energy, to overcome the spin-anisotropy energy and to rotate the moments
associated with antiferromagnetically correlated regions out of the $bc$
plane. To make a crude estimate, the correlation area of 4 $\times$ 10
nm$^2$ (Ref. \onlinecite{Matsuda}) contains roughly 300 Cu ions, for
which the integrated anisotropy energy for the WF moments, $\sum
zJ_{bc}SM_F/g\mu_B$ (where $z$ = 4 is the number of nearest neighbors,
$J_{bc} \simeq$ 0.7 meV is the in-plane anisotropic exchange, $S=1/2$ is
the Cu spin, and $M_F \simeq 0.002 \mu_B$ is the WF moment per Cu),
\cite{ThioFlop} is only about 0.4 meV --- this energy is actually
smaller than the Zeeman energy for the integrated WF moment in 40 T
($>1$ meV) and is comparable to the thermal energy at the spin-glass
ordering temperature (6--10 K), giving conficence in this estimate.
Therefore, it is likely that the spins are ``deconfined" from the $bc$
plane in high magnetic fields before the high-field WF state is
established, and this can explain the observed isotropy in the MR that
is expected to be associated with the reorientation of the WF moment.

Whatever the cause of the isotropy in the MR for $x$ = 0.03, it is
intriguing that the magnitude of the MR can become as large as 20\% at
low temperature, indicating that a large MR is a common feature of
insulating LSCO that shows some kind of magnetic order, with which the
WF moment can become well-defined. Thus, it is most reasonable to
conclude that the large MR in the insulating LSCO is essentially
associated with the reorientation of the WF moment and the resulting
change in the antiferromagnetic domain boundaries, into which the
doped-holes are presumably segregated. \cite{AndoMR}

\section{CONCLUSIONS}

We measure the magnetoresistance (MR) in the in-plane resistivity of
high-quality LSCO single crystals with $x$ = 0.01 and 0.03 in pulsed
magnetic fields up to 55 T. Contrary to the previous report by Thio {\it
et al.} in 1990, \cite{ThioFlop} we observe that there is no spin-flop
transition in the N\'eel state when the magnetic field is applied from
the spin easy axis ($b$ axis), and the high-field weak-ferromagnetic
(WF) state (where the WF moments are polarized along the $b$ axis) is
achieved in just one step. For $x$ = 0.03, a large (up to $\sim$20\%)
negative MR is observed in the spin-glass state, where the saturation of
this MR occurs at around 40 T. Intriguingly, the large MR in the
spin-glass state is essentially isotropic, which is probably related to
the short antiferromagnetic correlation length in the spin-glass state
that allows a deconfinement of spins out of the $bc$ planes in
high-magnetic fields.

\begin{acknowledgments}
We acknowledge fruitful discussions with A. Gozar and I. Tsukada. The
work at CRIEPI is supported in part by the Grant-in-Aid for Science
provided by the Japanese Society for the Promotion of Science. The NHMFL
is supported by the NSF, the State of Florida and the DOE.
\end{acknowledgments}



\begin{thebibliography}{99}
\bibitem{note}
In the variable-range-hopping regime, the ordinary MR is caused by a
magnetic-field-induced change in the distribution of the energy levels
and its sign is almost unpredictable.
\end{thebibliography}
\end{document}